\documentclass[final,12pt]{elsarticle}

\usepackage{amssymb}

\journal{International Journal of Pharmaceutics}

\begin{document}

\begin{frontmatter}



\title{On the dilemma of fractal or fractional kinetics in drug release studies:
	A comparison between Weibull and Mittag-Leffler functions}


\author[a,c]{Kosmas Kosmidis\corref{cor1}}
\author[b,c] {Panos Macheras} 

\address[a]{Physics Department, Aristotle University of Thessaloniki, 54124 Thessaloniki, Greece}
\address[b]{ Laboratory of Biopharmaceutics Pharmacokinetics, Department of Pharmacy,
	School of Health Sciences, National and Kapodistrian University of Athens,
	Athens, Greece}
\address[c]{ Pharmainformatics Unit “Athena” Research and Innovation Center, Athens,
	Greece
}

\cortext[cor1]{To whom correspondence should be addressed. (e-mail:kosmask@auth.gr)}

\begin{abstract}
We compare two of the most successful models for the description and analysis of drug release data. The fractal kinetics approach leading to release profiles described by a Weibull function and the fractional kinetics approach leading to release profiles described by a Mittag-Leffler function. We used Monte Carlo simulations to generate artificial release data from euclidean and fractal substrates. We have also used real release data from the literature and found that both models are capable in describing release data up to roughly $85\% $ of the release. For larger times both models systematically overestimate the number of particles remaining in the release device.
\end{abstract}


\end{frontmatter}


\section{Introduction}

Modeling the processes involved in controlled drug release is vital for the development of new pharmaceutical products\cite{PanosMacheras2016}. 
Although several models have been proposed for the description of drug release\cite{siepmann2001modeling,ritger1987simple,Savageau1995,Savageau1998,Marsh2006,Nygren1993,kosmidis2004mmk} , the main mathematical expressions used in pharmaceutics to describe the kinetics of drug release from a large variety of devices are
1. The Peppas equation or the so-called power law \cite{ritger1987simple}
\begin{equation}
\frac{M_t}{M_\infty}=k t^n
\label{eq:1}
\end{equation}
where $M_t$ and $M_\infty$ are the amounts of drug released at
times $t$ and infinity, respectively. In the above $k$ is an experimentally
determined parameter, and $n$ is an exponent that depends on the geometry of the boundary of the system which can be related to the drug release mechanisms. The power law model is probably the most commonly used model of drug release since it is rather easy to implement and is widely used to macroscopically classify the characteristics of the release kinetics. It is usually the starting point of any release study. 

2. The Weibull model, 
\begin{equation}
\frac{M_t}{M_\infty}=1-\exp(-a t^b)
\label{eq:2}
\end{equation}
where a , b are constants. This model has the form of a
stretched exponential and it is used in drug
release studies as well as in
dissolution studies. This functional form (as well as alternative forms based on it) is derived as an approximation in the framework of fractal kinetics \cite{Kopelman1988,kosmidis2003fkd,macheras2000heterogeneity,hadjitheodorou2013quantifying,hadjitheodorou2014analytical,christidi2016dynamics,villalobos2006monte,villalobos2017drug}.

Recently, an alternative way has been proposed to mathematically describe anomalous
diffusion. Diffusion in constrained and fractal topologies
is studied by the use of fractional calculus. Fractional calculus \cite{sokolov2002fractional,podlubny1998fractional, macheras2000heterogeneity,Dokoumetzidis2009} introduces derivatives and integrals of fractional order, such as half or 3 quarters. Differential equations with fractional derivatives can be used to describe anomalous kinetics without introducing
time-dependent coefficients as in fractal kinetics. Such equations have been shown to describe experimental data of anomalous diffusion more accurately \cite{sokolov2002fractional}. When applying fractional calculus to the classical kinetic models one practically replaces the usual derivatives with fractional ones. Thus, in the classical zero-order kinetics model after replacing the derivative of order 1 by a derivative of fractional order $a$ we derive the following equation:

\begin{equation}
\frac{{d^{a} X}}{{dt^{a} }} = k_{0}
\label{eq:zero-order}
\end{equation}
where $k_0$ is the kinetic constant.

The same process applied to first-order kinetics will lead to the so-called Mittag Lefler - Fractional kinetics model.
 
A fractional differential equation for the drug release problem gives rise to the expression \cite{Dokoumetzidis2009} 
\begin{equation}
\frac{M_t}{M_\infty}=1-E_a(-k_1 t^a)
\label{eq:3}
\end{equation}

where $k_1$ is a constant, and $ E_a (x)$ is the Mittag-
Leffler (ML) function of order $a$. This last function is a
generalization of the exponential function whose exact behavior is obtained for $a=1$.

In this paper we compare the fractal kinetics approach  leading to release profiles described by Eq. \ref{eq:2}  with the fractional kinetics approach leading to release profiles described by Eq. \ref{eq:3}. We compare the functions to each other, to their performance in describing the release from Monte Carlo simulation data of euclidean and fractal substrates and to their description of real data release profiles. We find that both functional forms are capable in describing release data up to $80-85\% $ of the release. Then they both systematically overestimate the number of particles remaining in the release device. From a purely practical point of view the Weibull model performs a little better than the Mittag-Lefler model and fitting the release curve to the data is considerably faster for the Weibull function compared to the ML function whose ``complex'' form (a large sum of Gamma functions is required) is computationally more expensive.

It should be noted that all the above described models are based on kinetic considerations. There is also the large class of , so called, mechanistic models that are very useful when a detailed description of the underlying physical processes is required. Mechanistic models use partial differential equations to quantify the mechanisms involved in drug release such as water transport in polymer tablets, swelling, drug diffusion and erosion and numerical methods to solve the resulting equations \cite{caccavo2014modeling,caccavo2015controlled,kaunisto2010mechanistic,lamberti2011controlled}.

\section{Methods}
\subsection*{Simulation of drug release from a Cylinder}
Following \cite{kosmidis2003rdr}, we assume here that the drug molecules move inside the cylinder by the mechanism of Fickian diffusion. Moreover, we assume
excluded volume interactions between the particles, meaning that each molecule occupies a volume V where no other molecule can be at the same time.
 We start with randomly distributed drug molecules and a known initial drug concentration inside the cylinder.
We first consider a three-dimensional cubic lattice with $L^3$ sites. We next define inside this cubic lattice a cylinder. A site is uniquely defined by its 3 coordinates $i,j,k$. If $r$ is the radius of the
cylinder and $i^2 + j^2 < (r - 1)^2$  then the site  belongs to the interior of the
cylinder and it can host drug molecules. If, on the other hand,
$i^2 + j^2 > r^2$ then it is outside the cylinder and it is a
restricted area, and particles are not allowed to go there. Finally, we label leak sites. We choose to label as leak sites the sites with indices $(r -1)^2 < i^2 + j^2 <r^2$ , thus defining a cylinder leaking from its round surface but
not from its top or bottom. Reflective boundary conditions are used for the top and bottom surface. Next, we place a number of particles randomly on the sites
of the cylinder, according to an initial particle concentration
$c=0.5$, avoiding double occupancy. This means
that $50\%$ of the sites are initially occupied by particles, and
the rest are empty. The diffusion process is simulated by selecting a particle at random and moving it to a randomly selected nearest neighbor site. If the new site is an empty site, then the move is allowed, and the particle is moved to this
new site. If the new site is already occupied, the move is
rejected. A particle is removed from the lattice as soon as it migrates to a
site lying within the leak area. After each particle move the time is incremented. The increment is chosen to
be $1/N$, where $N$ is the number of particles remaining in the
system. This is a typical approach in Monte Carlo simulations, and is necessary because the number of particles continuously decreases, and thus, the time unit characterizing the system is the mean time required for all $N$ particles present to move one step.
We average our results using different initial random configurations but the same parameters.

\subsection*{Simulation of drug release from a the percolation fractal}

Following \cite{kosmidis2003fkd} and references therein, we considered percolation fractals encapsulated on a square lattice at the percolation threshold $p_c=0.593$. The fractal dimension of the percolation fractal is known to be $91/48$. Calculations were performed as described below. 
For each run we generate a new fractal matrix assuming cyclic boundary conditions. We start with a known initial drug concentration $c=0.5$ and with randomly distributed drug molecules inside the fractal matrix. We again assume excluded volume interactions between the particles, meaning that two molecules cannot occupy the same site at the same time. The matrix can leak from the intersection of the percolation fractal with the boundaries of the square box where it is embedded.
The diffusion process is simulated in the same way as for the release from the cylinder. We monitor the
number of particles that are present inside the matrix as a
function of time until a fixed number of particles (50 particles) remains in the matrix. We average our results using different initial random configurations over at least 500 realizations.

\section{Results and Discussion}

\begin{figure} [t]
	\begin{center}
	\includegraphics[width=13.5cm]{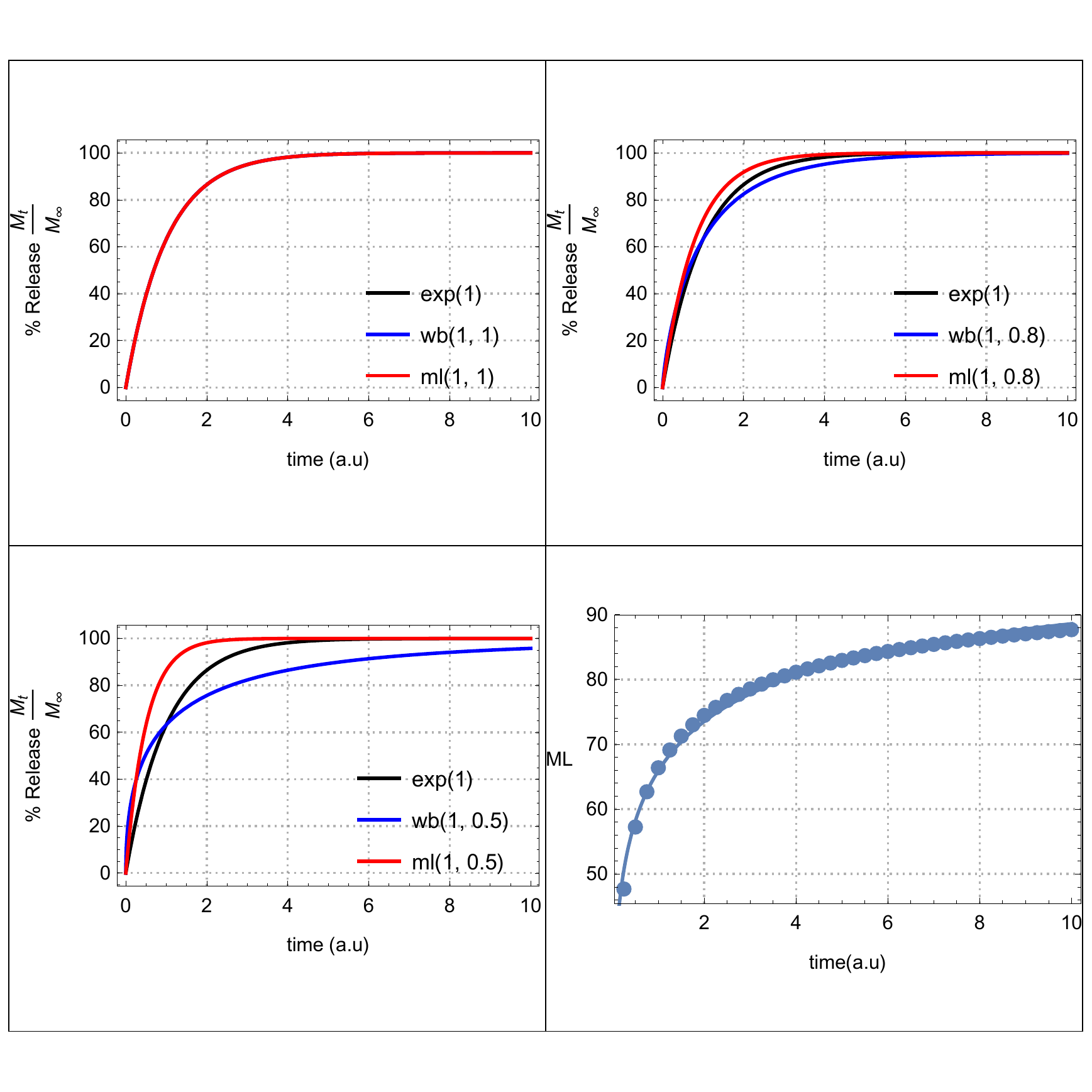}
	\caption{(color online) Comparison of the Weibull and Mittag-Lefler functions. An exponentially increasing function is shown as a black line for comparison in all figures. (Top Left) The blue line shows a Weibull function (wb) with $a=1$ and $b=1$ while the red a Mitage-Leffler function (ml) with $n1=1$ and $a=1$. The two functions are identical in this case and coincide with the exponential . (Top Right) The blue line shows a Weibull function (wb) with $a=1$ and $b=0.8$ while the red a Mittag-Leffler function (ml) with $n1=1$ and $a=0.8$. The two functions begin to diverge. (Bottom Left) The blue line shows a Weibull function (wb) with $a=1$ and $b=0.5$ while the red a Mittag-Leffler function (ml) with $n1=1$ and $a=0.5$. The two functions diverge more for decreasing values of the exponent. (Bottom Right) The points show a Mittag-Leffler function with $n1=1$ and $a=0.5$. The solid line is a fitting of a Weibull function with $a=1.08$ and $b=0.29$ to the points.Observe the descriptive power of the Weibull that leads to an almost ``perfect'' fitting. In all cases time is in arbitrary units (a.u)}
	\label{fig1}
	\end{center}
\end{figure}

In Fig.\ref{fig1} we compare the two basic functions i.e. the Weibull eq.\ref{eq:1} resulting from the fractal kinetics frameworks and the ML eq.\ref{eq:3} resulting from fractional calculus considerations.

 Figure\ref{fig1} (Top Left) shows a Weibull function (wb) with $a=1$ and $b=1$ (blue line) and a Mittag-Leffler function (ml) with $k_1=1$ and $a=1$ (red line) and an exponentially increasing function(black line). The three functions are identical in this case. Fig.\ref{fig1} (Top Right) shows with a blue line a Weibull function (wb) with $a=1$ and $b=0.8$, with black a Mittag-Leffler function (ml) with $k_1=1$ and $a=0.8$ and an exponentially increasing function(black line). We notice that the two functions  begin to diverge. In Fig. \ref{fig1} (Bottom Left) we present a Weibull function  with $a=1$ and $b=0.5$ (blue line), an ML function with $k_1=1$ and $a=0.5$ (red) and an exponential function(black line). We observe that the two functions diverge more for decreasing values of the exponent.In Fig\ref{fig1} (Bottom Right) we use points to show a Mittag-Leffler function with $n1=1$ and $a=0.5$. The solid line is a fitting of a Weibull function with $a=1.08$ and $b=0.29$ to the points. Observe the descriptive power of the Weibull that leads to an almost ``perfect'' fitting. This actually confirms that fractal and fractional kinetics considerations are both valid starting points for the description of the release problem and lead to equations that are rather close from a numerical point of view.
 
\begin{figure}[h]
	\begin{center}
	\includegraphics[width=12cm]{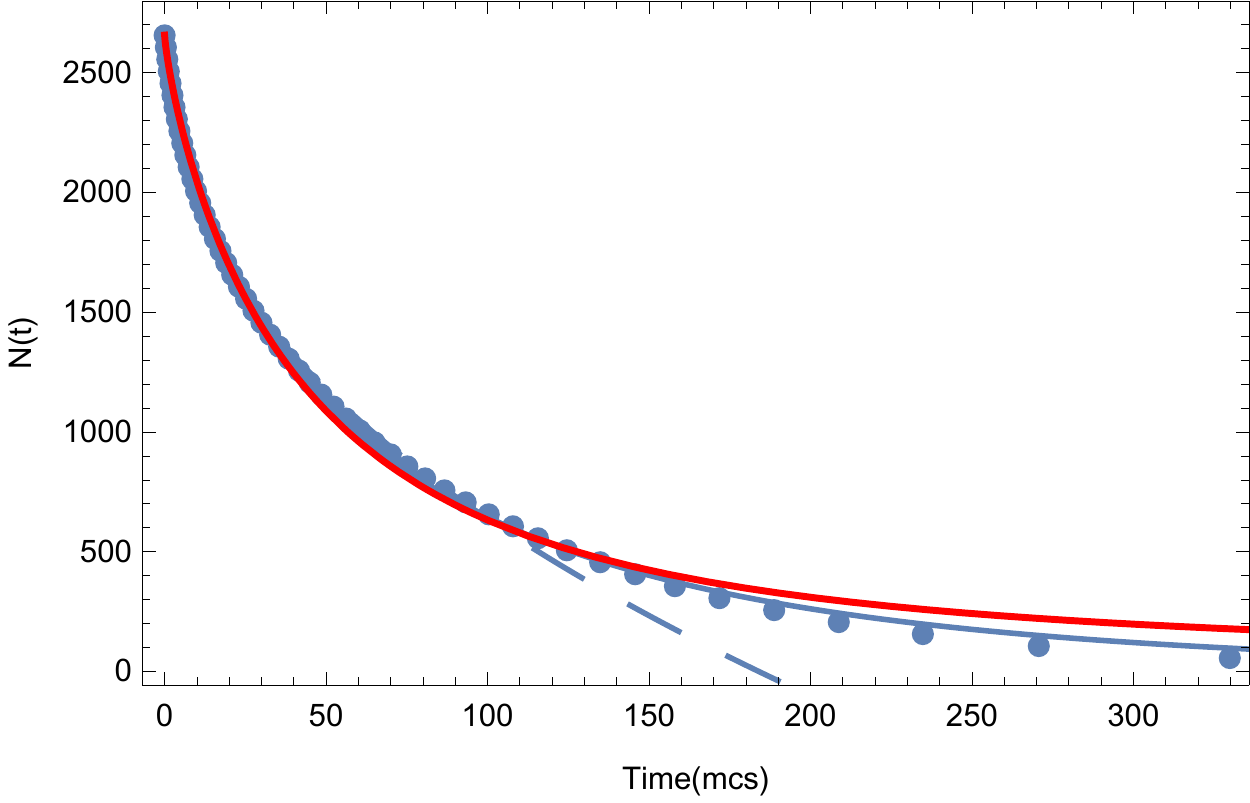}
	\caption{(color online) Monte Carlo simulations: Number of drug molecules $N(t)$ inside a cylinder as a function of time. Points are the Monte Carlo data. The dashed line is a fitting of the Peppas model $N(t)=N_0 (1-k t^0.45)$. The solid blue line is a fitting to a Weibull model $N(t)=N_0 \exp(-a t^b)$.Best fit obtained for $b=0.71$ . The solid red line is a fitting to a Mittag-Leffler model $N(t)=N_0 E_a(-k_1 t^a)$.Best fit obtained for $b=0.81$    }
	\label{fig2}
	\end{center}
\end{figure}

Figure \ref{fig2} shows Monte Carlo simulation results for the number of drug molecules $N(t)$ inside a cylinder as a function of time. The cylinder has a height 21 sites and a diameter of 21 sites. The initial number of drug molecules randomly distributed in the cylider is $N_0=2,657$. Points are the Monte Carlo data. The dashed line is a fitting of the Peppas model $N(t)=N_0 (1-k t^n)$ with $n=0.45$. The solid blue line is a fitting to a Weibull model $N(t)=N_0 \exp(-a t^b)$ . The solid red line is a fitting to a Mittag-Leffler model $N(t)=N_0 E_a(-k_1 t^a)$. Best fits were obtained for a Weibull function exponent $b=0.71$ and an ML exponent $a=0.81$. The Akaike Information criterion (AIC) and the Bayesian Information Criterion (BIC) have been calculated for each of the two functions. These two criteria are used to measure the goodness of fit of each model to the data. The model with the lower values of AIC and BIC is most probable to minimize information loss and thus signify a more suitable choice \cite{Burnham2002}. Here, for the  Mittag-Leffler model we have found $AIC=579.181, BIC= 585.148$ while for the  Weibull function $AIC=474.691, BIC= 480.658$.

\begin{figure}[h]
	\begin{center}
	\includegraphics[width=12cm]{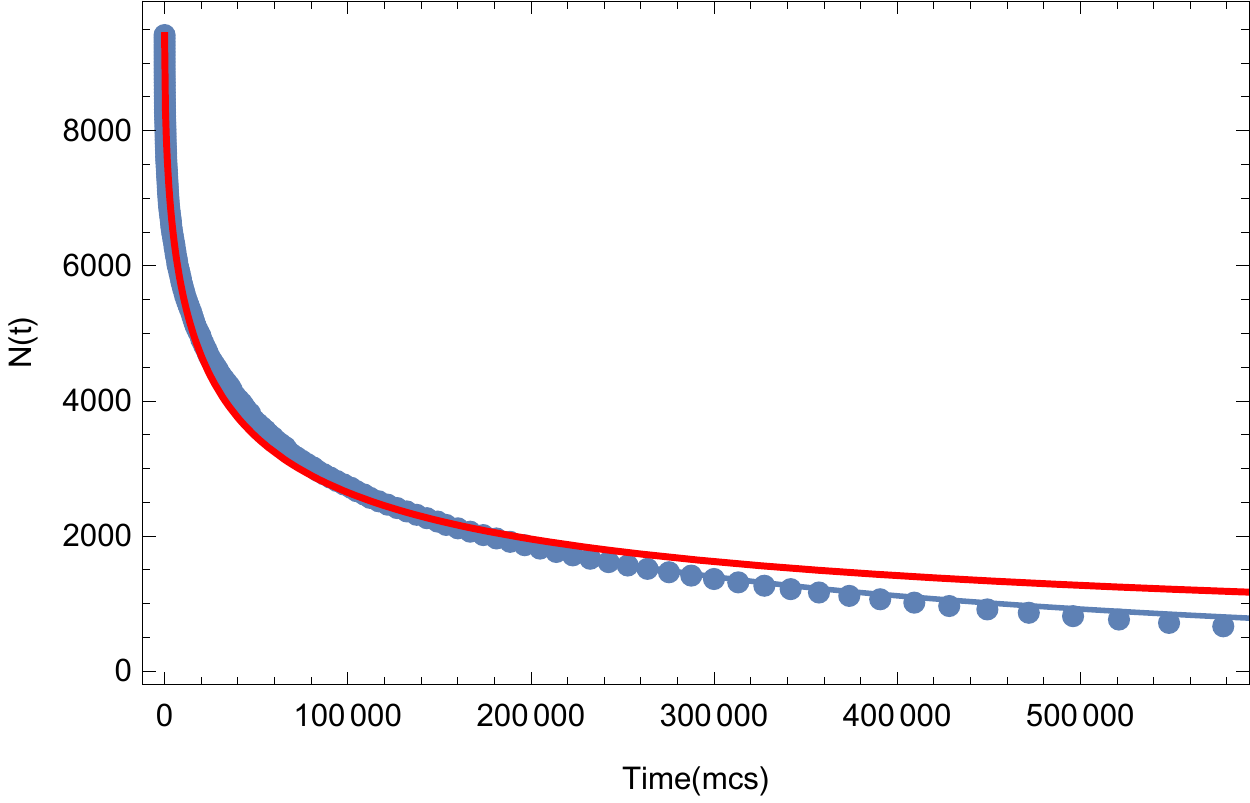}
	\caption{(color online) Monte Carlo simulations: Number of drug molecules $N(t)$ inside the percolation fractal as a function of time. Points are the Monte Carlo data.  The solid blue line is a fitting to a Weibull model $N(t)=N_0 \exp(-a t^b)$ . The solid red line is a fitting to a Mittag-Leffler model $N(t)=N_0 E_a(-k_1 t^a)$}
	\label{fig3}
	\end{center}
\end{figure}

Figure \ref{fig3} shows Monte Carlo simulation results for the number of drug molecules $N(t)$ inside percolation fractals embedded in $L=200$ square lattice as a function of time. Points are again the Monte Carlo data. The solid blue line is a fitting to a Weibull model $N(t)=N_0 \exp(-a t^b)$ . The solid red line is a fitting to a Mittag-Leffler model $N(t)=N_0 E_a(-k_1 t^a)$. Best fits were obtained for a Weibull function exponent $b=0.39$ and an ML exponent $a=0.51$. For the  Mittag-Leffler model we have found $AIC=2651.42, BIC= 2661.13$ while for the  Weibull function $AIC=2205.59, BIC= 2215.3$.

We observe that neither fractal nor fractional kinetics considerations can fully describe 100\% of the release from fractal matrices. In conditions of perfect ``mixing'' and a Euclidean space the release profile should be purely exponential. The departure from this exponential release in a Euclidean space is due to the creation of a depletion zone around the release sites. The existence of a disordered (fractal) environment is an additional reason for the departure from first order kinetics. The relative importance of the two mechanisms is not the same at the beginning of the release process as it is at the end where very few drug molecules remain and the depletion effect is not so strong. This fact is not taken care in deriving neither the Weibull nor the Mittag-Leffler approximation where mainly only the``imperfect'' mixing due the substrate disorder is taken into account.

\begin{figure}[h]
	\begin{center}
	\includegraphics[width=12cm]{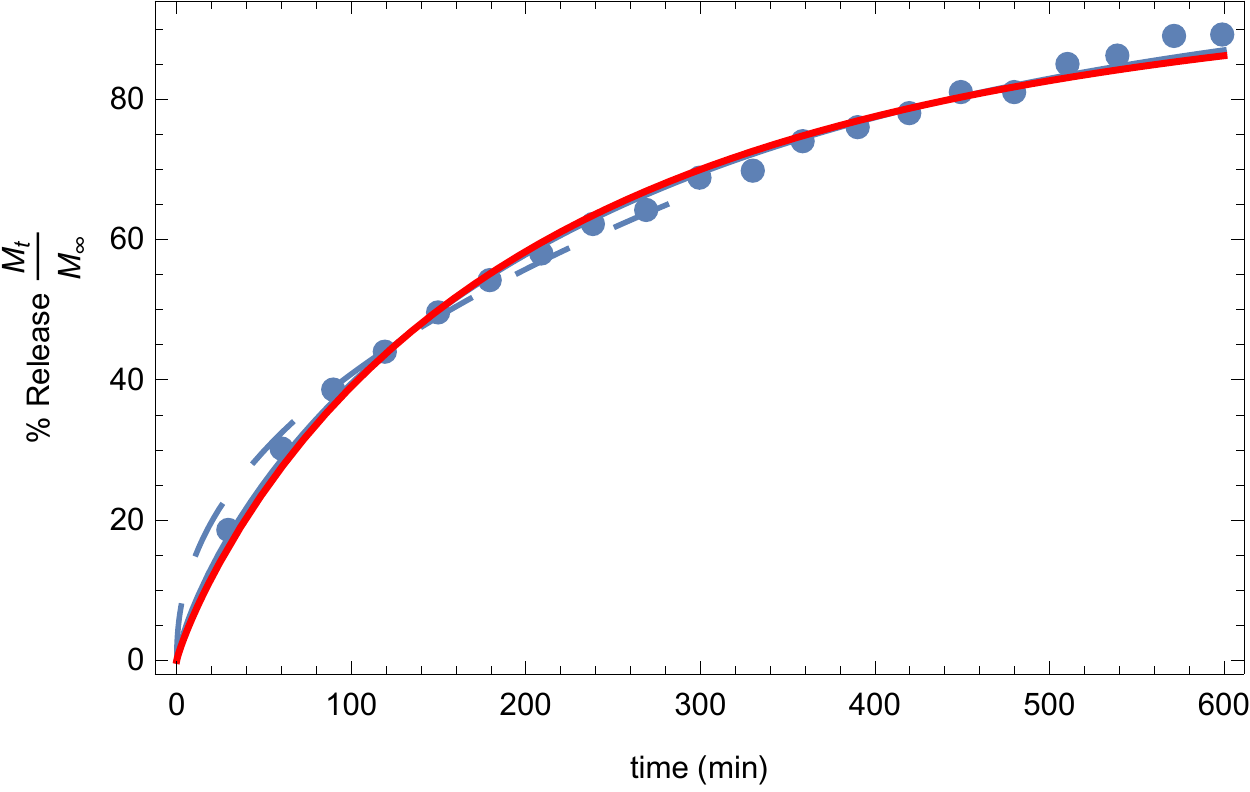}
	\caption{(color online) Fitting of the Peppas model Eq.\ref{eq:1} (dotted line), the Weibull model Eq.\ref{eq:2}(solid blue line) and the Mittag-Leffler model Eq.\ref{eq:3}(solid red line) to release data of Elvesil$^{\tiny{\textregistered }}$ (Papadopoulou et al., 2006, fig 1)}
	\label{fig4}
	\end{center}
\end{figure}
Next, we analyze drug release kinetics by plotting the mean
release data versus time from two characteristic cases from the literature. First we used data obtained from fig.1 of the publication \cite{papadopoulou2006uwf} concerning the release profile of tablets of Elvesil$^{\tiny{\textregistered }}$ 120mg diltiazem hydrochloride (Biomedica).
We studied the Elvesil$^{\tiny{\textregistered }}$ release profile under the framework of fractal and fractional kinetics. We present our results in fig. \ref{fig4}. We observe that both a Weibull function(blue line) with exponent $b=0.78$ as well as a Mittag-Leffler function(red line) with $a=0.87$ are capable of describing the release data. The Power law model is shown (dashed line) for comparison.  For the  Mittag-Leffler model we have found $AIC=88.9639, BIC=91.9511$ while for the  Weibull function $AIC=76.8052, BIC=79.7924$.

\begin{figure}
	\begin{center}
	\includegraphics[width=12cm]{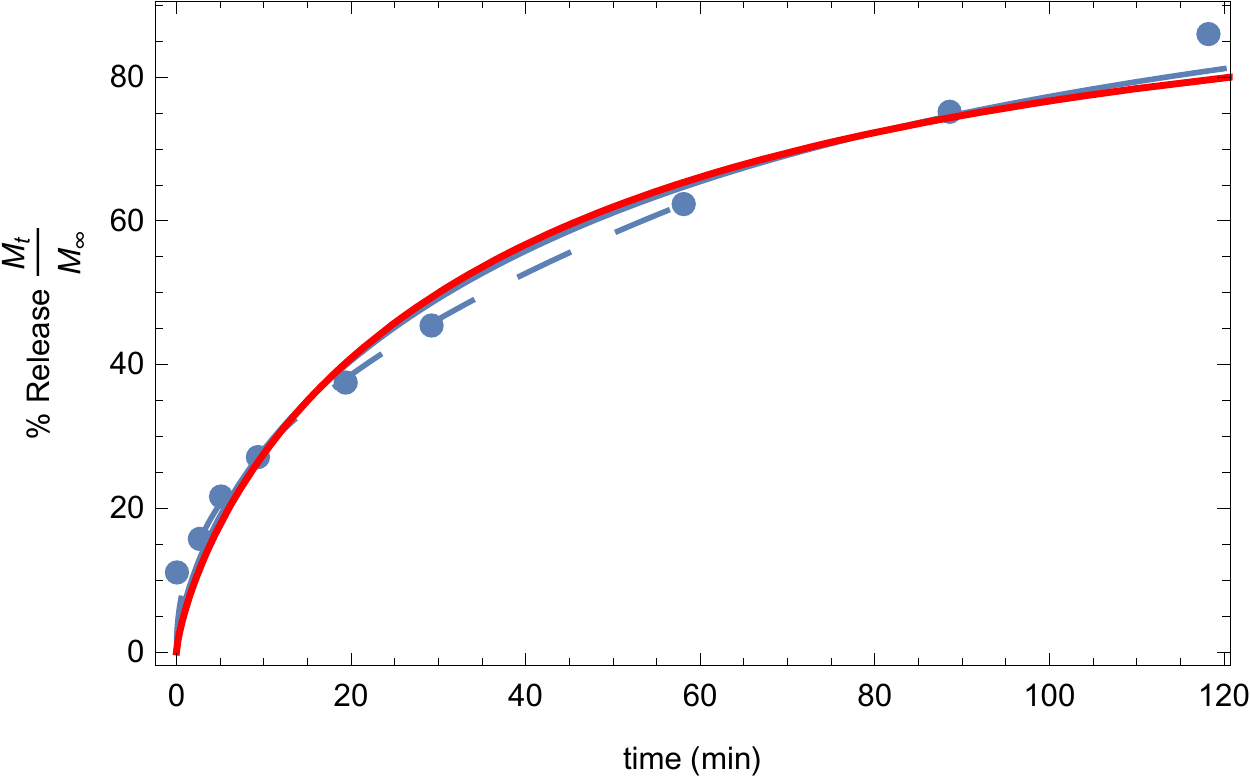}
	\caption{(color online) Fitting of the Peppas model Eq.\ref{eq:1} (dotted line), the Weibull model Eq.\ref{eq:2}(solid blue line) and the Mittag-Leffler model Eq.\ref{eq:3}(solid red line)  to literature release
		data of 4-aminopyridine/Metolose 90SH-4000SR (Juarez et al., 2001, Fig. 5)}
	\label{fig5}
	\end{center}
\end{figure}
Finally, we analyze data of the drug release profile of 4-aminopyridine/Metolose 90SH-4000SR matrices using HCl 0.1 N and phosphate buffer pH 7.4 as dissolution medium. Data were obtained  from fig.5 of the publication \cite{juarez2001influence}. We present our results in figure \ref{fig5}. We again observe that both a Weibull function(blue line) with exponent$b=0.65$ as well as a Mittag-Leffler function(red line) with $a=0.75$ are capable of describing the release data.  The Power law model is shown (dashed line) for comparison.  For the  Mittag-Leffler model we have found $AIC=60.2217, BIC= 60.8134$ while for the  Weibull function $AIC=57.7005, BIC= 58.2922$.
Both functions tend to underestimate the release ratio above $80\%$ of the release data in this case. The small value of the Weibull exponent is indicative of release from a rather disordered substrate. 

\section{Conclusions}

In this paper we compared the fractal kinetics approach  leading to release profiles described by Eq. \ref{eq:2}  with the fractional kinetics approach leading to release profiles described by Eq.\ref{eq:3}. We used Monte Carlo simulation data of euclidean and fractal substrates as well as real release data from the literature and found that both functional forms are capable in describing release data up to approximately $85\% $ of the release. Then they both systematically overestimate the number of particles remaining in the release device. From a purely practical point of view the Weibull model performs a little better than the Mittag-Leffler model concerning computational time.At the very end of the simulations runs, the descriptive (fitting) ability of the two functions differ considerably. However, one should take into
account the heterogeneous distribution of the particles created at the end
of the simulation period \cite{kosmidis2003fkd}. It seems likely that the
Weibull function, which relies on a time dependent coefficient \cite{macheras2000heterogeneity} deviates less from the simulation pattern than the ML function,
which is based on a differential equation of non integer order. Overall,
both functions deviate from the late stage of the simulation pattern;
however, the Weibull function exhibits smaller deviation than the ML
function. This can be seen from the visual inspection of figures \ref{fig2}-\ref{fig5} and from the fact that the AIC and BIC estimations for the Weibull model are in all cases smaller than that of the Mittag-Leffler model.  As far as the fitting to experimental data is concerned the
Weibull function exhibits again better performance on the basis of
classical statistical criteria (correlation coefficient, distribution of
residuals , AIC) than the ML function. In short, the Weibul and the Mittag-Leffler model are practically equally good in describing drug release profiles. They are not reliable for the description of the final stage of the release process and in such cases both should be used with caution. From a purely practical point of view the Weibull model is easier to use.





\bibliographystyle{model1-num-names}
\bibliography{Fractal_kinetics}

\begin{thebibliography}{27}
\expandafter\ifx\csname natexlab\endcsname\relax\def\natexlab#1{#1}\fi
\providecommand{\bibinfo}[2]{#2}
\ifx\xfnm\relax \def\xfnm[#1]{\unskip,\space#1}\fi
\bibitem[{Macheras and Iliadis(2016)}]{PanosMacheras2016}
\bibinfo{author}{P.~Macheras}, \bibinfo{author}{A.~Iliadis},
  \bibinfo{title}{Modeling in Biopharmaceutics, Pharmacokinetics and
  Pharmacodynamics}, \bibinfo{publisher}{Springer International Springer
  International Publishing}, \bibinfo{year}{2016}.
\bibitem[{Siepmann and Peppas(2001)}]{siepmann2001modeling}
\bibinfo{author}{J.~Siepmann}, \bibinfo{author}{N.~Peppas},
\newblock \bibinfo{title}{Modeling of drug release from delivery systems based
  on hydroxypropyl methylcellulose (hpmc)},
\newblock \bibinfo{journal}{Advanced drug delivery reviews}
  \bibinfo{volume}{48} (\bibinfo{year}{2001}) \bibinfo{pages}{139--157}.
\bibitem[{Ritger and Peppas(1987)}]{ritger1987simple}
\bibinfo{author}{P.~L. Ritger}, \bibinfo{author}{N.~A. Peppas},
\newblock \bibinfo{title}{A simple equation for description of solute release
  ii. fickian and anomalous release from swellable devices},
\newblock \bibinfo{journal}{Journal of controlled release} \bibinfo{volume}{5}
  (\bibinfo{year}{1987}) \bibinfo{pages}{37--42}.
\bibitem[{Savageau(1995)}]{Savageau1995}
\bibinfo{author}{M.~A. Savageau},
\newblock \bibinfo{title}{Michaelis-menten mechanism reconsidered: implications
  of fractal kinetics.},
\newblock \bibinfo{journal}{J Theor Biol} \bibinfo{volume}{176}
  (\bibinfo{year}{1995}) \bibinfo{pages}{115--124}.
\bibitem[{Savageau(1998)}]{Savageau1998}
\bibinfo{author}{M.~A. Savageau},
\newblock \bibinfo{title}{Development of fractal kinetic theory for
  enzyme-catalysed reactions and implications for the design of biochemical
  pathways.},
\newblock \bibinfo{journal}{Biosystems} \bibinfo{volume}{47}
  (\bibinfo{year}{1998}) \bibinfo{pages}{9--36}.
\bibitem[{Marsh and Tuszyński(2006)}]{Marsh2006}
\bibinfo{author}{R.~E. Marsh}, \bibinfo{author}{J.~A. Tuszyński},
\newblock \bibinfo{title}{Fractal michaelis-menten kinetics under steady state
  conditions: Application to mibefradil.},
\newblock \bibinfo{journal}{Pharm Res} \bibinfo{volume}{23}
  (\bibinfo{year}{2006}) \bibinfo{pages}{2760--2767}.
\bibitem[{Nygren(1993)}]{Nygren1993}
\bibinfo{author}{H.~Nygren},
\newblock \bibinfo{title}{Nonlinear kinetics of ferritin adsorption.},
\newblock \bibinfo{journal}{Biophys J} \bibinfo{volume}{65}
  (\bibinfo{year}{1993}) \bibinfo{pages}{1508--1512}.
\bibitem[{Kosmidis et~al.(2004)Kosmidis, Karalis, Argyrakis, and
  Macheras}]{kosmidis2004mmk}
\bibinfo{author}{K.~Kosmidis}, \bibinfo{author}{V.~Karalis},
  \bibinfo{author}{P.~Argyrakis}, \bibinfo{author}{P.~Macheras},
\newblock \bibinfo{title}{Michaelis-menten kinetics under spatially constrained
  conditions: Application to mibefradil pharmacokinetics},
\newblock \bibinfo{journal}{Biophysical Journal} \bibinfo{volume}{87}
  (\bibinfo{year}{2004}) \bibinfo{pages}{1498--1506}.
\bibitem[{Kopelman(1988)}]{Kopelman1988}
\bibinfo{author}{R.~Kopelman},
\newblock \bibinfo{title}{Fractal reaction kinetics},
\newblock \bibinfo{journal}{Science} \bibinfo{volume}{241, No. 4873}
  (\bibinfo{year}{1988}) \bibinfo{pages}{1620--1626}.
\bibitem[{Kosmidis et~al.(2003)Kosmidis, Argyrakis, and
  Macheras}]{kosmidis2003fkd}
\bibinfo{author}{K.~Kosmidis}, \bibinfo{author}{P.~Argyrakis},
  \bibinfo{author}{P.~Macheras},
\newblock \bibinfo{title}{Fractal kinetics in drug release from finite fractal
  matrices},
\newblock \bibinfo{journal}{The Journal of Chemical Physics}
  \bibinfo{volume}{119} (\bibinfo{year}{2003}) \bibinfo{pages}{6373}.
\bibitem[{Macheras and Dokoumetzidis(2000)}]{macheras2000heterogeneity}
\bibinfo{author}{P.~Macheras}, \bibinfo{author}{A.~Dokoumetzidis},
\newblock \bibinfo{title}{On the heterogeneity of drug dissolution and
  release},
\newblock \bibinfo{journal}{Pharmaceutical research} \bibinfo{volume}{17}
  (\bibinfo{year}{2000}) \bibinfo{pages}{108--112}.
\bibitem[{Hadjitheodorou and Kalosakas(2013)}]{hadjitheodorou2013quantifying}
\bibinfo{author}{A.~Hadjitheodorou}, \bibinfo{author}{G.~Kalosakas},
\newblock \bibinfo{title}{Quantifying diffusion-controlled drug release from
  spherical devices using monte carlo simulations},
\newblock \bibinfo{journal}{Materials Science and Engineering: C}
  \bibinfo{volume}{33} (\bibinfo{year}{2013}) \bibinfo{pages}{763--768}.
\bibitem[{Hadjitheodorou and Kalosakas(2014)}]{hadjitheodorou2014analytical}
\bibinfo{author}{A.~Hadjitheodorou}, \bibinfo{author}{G.~Kalosakas},
\newblock \bibinfo{title}{Analytical and numerical study of
  diffusion-controlled drug release from composite spherical matrices},
\newblock \bibinfo{journal}{Materials Science and Engineering: C}
  \bibinfo{volume}{42} (\bibinfo{year}{2014}) \bibinfo{pages}{681--690}.
\bibitem[{Christidi and Kalosakas(2016)}]{christidi2016dynamics}
\bibinfo{author}{E.~Christidi}, \bibinfo{author}{G.~Kalosakas},
\newblock \bibinfo{title}{Dynamics of the fraction of drug particles near the
  release boundary},
\newblock \bibinfo{journal}{The European Physical Journal Special Topics}
  \bibinfo{volume}{225} (\bibinfo{year}{2016}) \bibinfo{pages}{1245--1254}.
\bibitem[{Villalobos et~al.(2006)Villalobos, Vidales, Cordero, Quintanar, and
  Dom{\'\i}nguez}]{villalobos2006monte}
\bibinfo{author}{R.~Villalobos}, \bibinfo{author}{A.~M. Vidales},
  \bibinfo{author}{S.~Cordero}, \bibinfo{author}{D.~Quintanar},
  \bibinfo{author}{A.~Dom{\'\i}nguez},
\newblock \bibinfo{title}{Monte carlo simulation of diffusion-limited drug
  release from finite fractal matrices},
\newblock \bibinfo{journal}{Journal of sol-gel science and technology}
  \bibinfo{volume}{37} (\bibinfo{year}{2006}) \bibinfo{pages}{195--199}.
\bibitem[{Villalobos et~al.(2017)Villalobos, V~Garcia, Quintanar, and
  M~Young}]{villalobos2017drug}
\bibinfo{author}{R.~Villalobos}, \bibinfo{author}{E.~V~Garcia},
  \bibinfo{author}{D.~Quintanar}, \bibinfo{author}{P.~M~Young},
\newblock \bibinfo{title}{Drug release from inert spherical matrix systems
  using monte carlo simulations},
\newblock \bibinfo{journal}{Current drug delivery} \bibinfo{volume}{14}
  (\bibinfo{year}{2017}) \bibinfo{pages}{65--72}.
\bibitem[{Sokolov et~al.(2002)Sokolov, Klafter, and
  Blumen}]{sokolov2002fractional}
\bibinfo{author}{I.~M. Sokolov}, \bibinfo{author}{J.~Klafter},
  \bibinfo{author}{A.~Blumen},
\newblock \bibinfo{title}{Fractional kinetics},
\newblock \bibinfo{journal}{Physics Today} \bibinfo{volume}{55}
  (\bibinfo{year}{2002}) \bibinfo{pages}{48--54}.
\bibitem[{Podlubny(1998)}]{podlubny1998fractional}
\bibinfo{author}{I.~Podlubny}, \bibinfo{title}{Fractional differential
  equations: an introduction to fractional derivatives, fractional differential
  equations, to methods of their solution and some of their applications},
  volume \bibinfo{volume}{198}, \bibinfo{publisher}{Elsevier},
  \bibinfo{year}{1998}.
\bibitem[{Dokoumetzidis and Macheras(2009)}]{Dokoumetzidis2009}
\bibinfo{author}{A.~Dokoumetzidis}, \bibinfo{author}{P.~Macheras},
\newblock \bibinfo{title}{Fractional kinetics in drug absorption and
  disposition processes.},
\newblock \bibinfo{journal}{J Pharmacokinet Pharmacodyn} \bibinfo{volume}{36}
  (\bibinfo{year}{2009}) \bibinfo{pages}{165--178}.
\bibitem[{Caccavo et~al.(2014)Caccavo, Cascone, Lamberti, and
  Barba}]{caccavo2014modeling}
\bibinfo{author}{D.~Caccavo}, \bibinfo{author}{S.~Cascone},
  \bibinfo{author}{G.~Lamberti}, \bibinfo{author}{A.~A. Barba},
\newblock \bibinfo{title}{Modeling the drug release from hydrogel-based
  matrices},
\newblock \bibinfo{journal}{Molecular pharmaceutics} \bibinfo{volume}{12}
  (\bibinfo{year}{2014}) \bibinfo{pages}{474--483}.
\bibitem[{Caccavo et~al.(2015)Caccavo, Cascone, Lamberti, and
  Barba}]{caccavo2015controlled}
\bibinfo{author}{D.~Caccavo}, \bibinfo{author}{S.~Cascone},
  \bibinfo{author}{G.~Lamberti}, \bibinfo{author}{A.~A. Barba},
\newblock \bibinfo{title}{Controlled drug release from hydrogel-based matrices:
  Experiments and modeling},
\newblock \bibinfo{journal}{International journal of pharmaceutics}
  \bibinfo{volume}{486} (\bibinfo{year}{2015}) \bibinfo{pages}{144--152}.
\bibitem[{Kaunisto et~al.(2010)Kaunisto, Abrahmsen-Alami, Borgquist, Larsson,
  Nilsson, and Axelsson}]{kaunisto2010mechanistic}
\bibinfo{author}{E.~Kaunisto}, \bibinfo{author}{S.~Abrahmsen-Alami},
  \bibinfo{author}{P.~Borgquist}, \bibinfo{author}{A.~Larsson},
  \bibinfo{author}{B.~Nilsson}, \bibinfo{author}{A.~Axelsson},
\newblock \bibinfo{title}{A mechanistic modelling approach to polymer
  dissolution using magnetic resonance microimaging},
\newblock \bibinfo{journal}{Journal of Controlled Release}
  \bibinfo{volume}{147} (\bibinfo{year}{2010}) \bibinfo{pages}{232--241}.
\bibitem[{Lamberti et~al.(2011)Lamberti, Galdi, and
  Barba}]{lamberti2011controlled}
\bibinfo{author}{G.~Lamberti}, \bibinfo{author}{I.~Galdi},
  \bibinfo{author}{A.~A. Barba},
\newblock \bibinfo{title}{Controlled release from hydrogel-based solid
  matrices. a model accounting for water up-take, swelling and erosion},
\newblock \bibinfo{journal}{International journal of pharmaceutics}
  \bibinfo{volume}{407} (\bibinfo{year}{2011}) \bibinfo{pages}{78--86}.
\bibitem[{Kosmidis et~al.(2003)Kosmidis, Argyrakis, and
  Macheras}]{kosmidis2003rdr}
\bibinfo{author}{K.~Kosmidis}, \bibinfo{author}{P.~Argyrakis},
  \bibinfo{author}{P.~Macheras},
\newblock \bibinfo{title}{A reappraisal of drug release laws using monte carlo
  simulations: The prevalence of the weibull function},
\newblock \bibinfo{journal}{Pharmaceutical Research} \bibinfo{volume}{20}
  (\bibinfo{year}{2003}) \bibinfo{pages}{988--995}.
\bibitem[{Burnham(2002)}]{Burnham2002}
\bibinfo{author}{D.~R. Burnham, K. P.;~Anderson}, \bibinfo{title}{, Model
  Selection and Multimodel Inference: A practical information-theoretic
  approach (2nd ed.)}, \bibinfo{publisher}{Springer-Verlag},
  \bibinfo{year}{2002}.
\bibitem[{Papadopoulou et~al.(2006)Papadopoulou, Kosmidis, Vlachou, and
  Macheras}]{papadopoulou2006uwf}
\bibinfo{author}{V.~Papadopoulou}, \bibinfo{author}{K.~Kosmidis},
  \bibinfo{author}{M.~Vlachou}, \bibinfo{author}{P.~Macheras},
\newblock \bibinfo{title}{On the use of the weibull function for the
  discernment of drug release mechanisms},
\newblock \bibinfo{journal}{International Journal of Pharmaceutics}
  \bibinfo{volume}{309} (\bibinfo{year}{2006}) \bibinfo{pages}{44--50}.
\bibitem[{Ju{\'a}rez et~al.(2001)Ju{\'a}rez, Rico, and
  Villafuerte}]{juarez2001influence}
\bibinfo{author}{H.~Ju{\'a}rez}, \bibinfo{author}{G.~Rico},
  \bibinfo{author}{L.~Villafuerte},
\newblock \bibinfo{title}{Influence of admixed carboxymethylcellulose on
  release of 4-aminopyridine from hydroxypropyl methylcellulose matrix
  tablets},
\newblock \bibinfo{journal}{International journal of pharmaceutics}
  \bibinfo{volume}{216} (\bibinfo{year}{2001}) \bibinfo{pages}{115--125}.

\end{thebibliography}







\end{document}